\documentclass[10pt,english,prl,showpacs,superscriptaddress]{revtex4}
\usepackage[latin9]{inputenc}
\usepackage{textcomp}
\usepackage{amsmath}
\usepackage{amssymb}
\usepackage{graphicx}

\makeatletter
\@ifundefined{textcolor}{}
{%
 \definecolor{BLACK}{gray}{0}
 \definecolor{WHITE}{gray}{1}
 \definecolor{RED}{rgb}{1,0,0}
 \definecolor{GREEN}{rgb}{0,1,0}
 \definecolor{BLUE}{rgb}{0,0,1}
 \definecolor{CYAN}{cmyk}{1,0,0,0}
 \definecolor{MAGENTA}{cmyk}{0,1,0,0}
 \definecolor{YELLOW}{cmyk}{0,0,1,0}
}

\usepackage{epstopdf}
\usepackage{color}
\makeatother

\usepackage{babel}
\begin{document}

\title{Temporal localized structures in photonic crystal fiber resonators and their spontaneous symmetry breaking instability}

\author{
L. Bahloul$^{1}$, L. Cherbi$^{1}$,  A. Hariz$^{1}$ and M. Tlidi$^{2}$}

\address{Laboratoire d'instrumentation, Universit{\'{e}} des Sciences et de la Technologie Houari Boumediene (USTHB), Alg\'{e}rie$^{1}$\\
Universit\'{e} Libre de Bruxelles (U.L.B.), Facult{\'{e}} des Sciences,  CP. 231, Campus Plaine, B-1050 Bruxelles, Belgium$^{2}$
}

\begin{abstract}
We investigate analytically and numerically the formation of temporal localized structures in all  photonic crystal fiber resonator. These dissipative structures  consist of isolated or randomly distributed peaks in an uniform background of the intensity profile. The number of peaks and their temporal distribution are determined solely by initial conditions. They exhibit multistability behavior in a finite range of parameters.  A weakly nonlinear analysis is performed in the neighborhood of the first threshold associated with the modulational instability. We consider the regime where the instability is not degenerate. We show that the fourth order dispersion affects the threshold associated with the formation of bright temporal localized structures. We estimate analytically and numerically both the linear and the nonlinear correction to the velocity of moving temporal structures induced by a spontaneous broken reflection symmetry mediated by the third-order dispersion. Finally, we show  that the third order dispersion affects the threshold associated with the moving temporal localized structures.
\end{abstract}


\maketitle

\section{Introduction}
Driven all fiber cavities constitute a basic configuration in nonlinear fiber optics.  More specifically, experimental studies have demonstrated that when these cavities are pumped by a continuous wave, they exhibit   spontaneously self-organized temporal structures in the form of trains of short pulses with well defined repetition rate \cite{Mitschke,COEN-Hal}. The theoretical prediction of this phenomenon was carried out in the seminal paper by Lugiato and Lefever  (LL model, \cite{LL}). The breakup of continuous wave into trains of pulses is attributed to the competition between the following phenomena (i) a nonlinear mechanism which is originated from the intensity-dependent refractive index  (Kerr effect) that tends to amplify locally the field intensity, (ii) a chromatic dispersion which on the contrary tends to restore uniformity, and (iii) dissipation.

Besides a train pulses distribution, temporal localized structures (TLS) are found in a well-defined region of parameters called a pinning zone. In this regime, the system exhibits a coexistence between two states: the uniform background and the train of pulses of light that emerges from subcritical modulational instability \cite{Scroggie}.  In the same way as in the temporal regime where the  breakup of continuous wave into trains of  pulses results from the interplay between nonlinearity and group velocity dispersion, in spatial cavities, the competition between nonlinearity and diffraction induces the formation of two-dimensional localized structures \cite{Scroggie,TML94,TOL2000,VLT11}. When the dispersion and diffraction have a comparable influence, their competition with nonlinearity could produce  varieties of three-dimensional periodic and localized "light bullet patterns" \cite{TLIDI2D,staliunas-3D,Tlidi-3D,Brambilla,tassin,chao}. These structures consist of regular 3D lattices of bright light bullet traveling at the group velocity of light in the material. 

Temporal localized structures are often called dissipative solitons or cavity solitons.  Experimental observations of TLS in all fiber nonlinear cavity has stimulated further the interest in this field  \cite{LEO}. Nowadays, temporal localized structures in standard silica optical fibers is an active field due to the maturity of the fiber technology and the possible applications as an ideal support for bits in an optical buffer that could be used for all-optical storage, all-optical reshaping and wavelength conversion \cite{LEO}. Recently, experimental study reveals that the interaction between two temporal localized structures is ultra-weak \cite{Jang}. For large intensity regime, temporal cavity solitons could exhibit a self-pulsation or chaotic behavior \cite{Gomila2005,Turaev,LEO-OE,Egorov}. Front propagation and switching waves between the two stable homogeneous steady states has been investigated experimentally and theoretically  in nonlinear all fiber cavities subject to injection \cite{Ceon-Tlidi}. More recently, the study of front propagation into an unstable state reveals that during time evolution, the velocity of propagating front evolves according to the universal power law \cite{Coulibaly}.

In all above mentioned studies, the dispersion  is limited to fiber cavities with
group-velocity dispersion restricted to the second order. However, when an optical cavity is operating
close to the zero dispersion wavelength, high-order chromatic dispersion effects could play an important role in the dynamics of  photonic crystal fiber (PCF) \cite{Cavalcanti91,Pitois03-2,joly05}. 
Photonic crystal fiber permit a high control of the dispersion curve and allow exploring previously inaccessible parameter regimes \cite{Russel,Schmidberger12}. The inclusion of the fourth order dispersion in the description of all fiber cavities,   permits the modulational instability to have a finite domain of existence delimited by two pump power values \cite{Tlidi1} and  allows for the stabilization of dark temporal localized structures \cite{Tlidi-Gelens}. Recently, it has been shown that the  combined influence of the third- and the fourth-order dispersion induced a motion of dark localized structures in large intensity regime \cite{Tlidi-Lyes}. These moving solutions involved an asymmetric odd or even number of dips which coexist for finite values of the input field intensity. When the spatio-temporal dynamics of PCFs cavities is ruled by ultrashort pulses, mutlistability as well as a spontaneous breaking in pulse-shape symmetry has been observed  \cite{Nicolas-OE-14}. More recently, the role of the third order dispersion in the dynamics of the bright temporal localized structures in micro-ring resonators pumped in a proximity of the zero of the group velocity dispersion has been reported \cite{Skryanbin-14}.

In this paper, it is our aim to analyze analytically the influence of high order dispersion on the nonlinear dynamical properties of bright temporal localized structures in all photonic crystal fiber cavity. We derive a normal form in the vicinity of the first modulational instability threshold by taking into account of the second, the third and the fourth order dispersion. The weakly nonlinear analysis allows us to determine the threshold associated with the emergence of TLS. In addition, we show that the third order dispersion corrects the linear velocity and pushes TLS to move with a constant velocity. This regular drift is induced by a broken reflection symmetry mediated by a third-order dispersion. 

The paper is organized as follows, after briefly introducing the model of photonic crystal fiber cavity, we provide  a summary of a linear stability analysis of the homogeneous steady states in Sec. 2. We present a weakly nonlinear analysis and the estimation of the nonlinear velocity associated with the moving periodic temporal structures in Sec. 3. Stationary and moving temporal localized structures are studied in Sec. 4. We conclude in sec. 5.

\begin{figure}[bbp]
\begin{center}
\includegraphics[width=18cm,height=12cm]{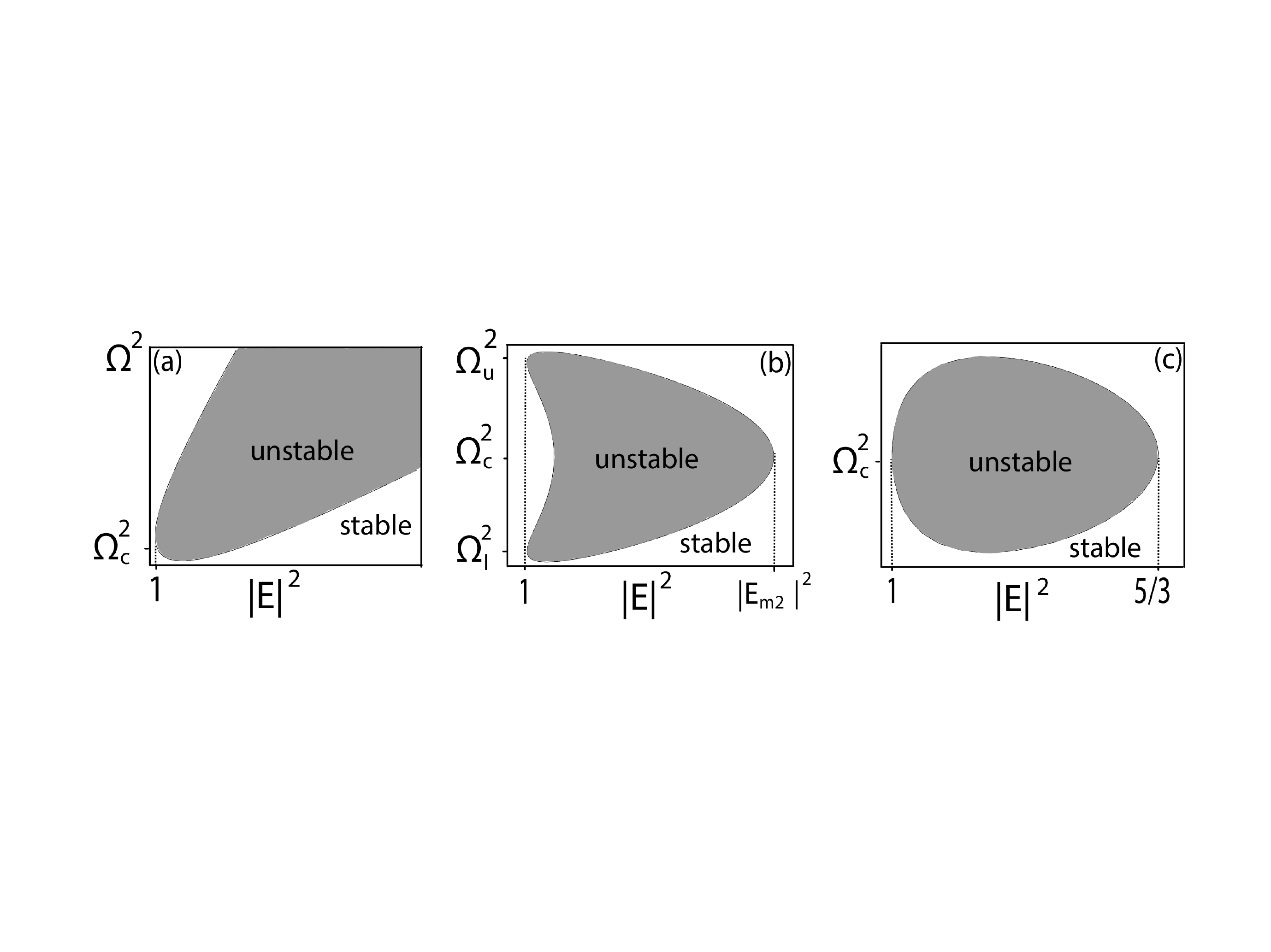}
\end{center}
\caption{Marginal stability curves  for homogenous steady states representing the instability regions in the plane $(|E|^2,\Omega^2$) for a fixed value of the detuning parameter $\Delta=1.3$.  (a) $B_4=0$ and  $B_2=-1.$ (b) $B_4=0.5$ and $B_2=-1.5$
(c) $B_4=0.5$ and $B_2=-1.1832$}
\label{Fig1}
\end{figure} 

\section{Model equation}
We consider a single mode photonic crystal fiber cavity pumped by a continuous wave of power $S^2$. Propagation of light inside the fiber is governed by the nonlinear Schr{\"o}dinger equation  \cite{Agrawal}, the use of PCFs allows to expand the propagation constant up to the fourth order in a Taylor series. This equation  is supplemented by appropriate resonator boundary conditions. The nonlinear Schr{\"o}dinger equation combined with boundary conditions leads to the equation that characterizes the propagation of light along the cavity, which is described by the generalized Lugiato-Lefever model with non dimensional variables as \cite{Tlidi1}: 
\begin{eqnarray}\frac{\partial E}{\partial t} &=&S-(1+i\Delta )E +i\left|
E \right| ^{2}E  \nonumber \\
&-&iB_{2}\frac{\partial ^{2}E }{\partial \tau
^{2}} +B_{3}\frac{\partial ^{3}E }{\partial \tau ^{3}}+iB_{4}\frac{%
\partial ^{4}E }{\partial \tau ^{ 4}}  \label{eq:E}
\end{eqnarray}
Where $E$ is the slowly varying envelope of the electric field propagating inside the cavity. The time $t$ is the slow scale time that describes the evolution of the field envelope $E$ from one cavity round trip to the other. The coefficients $B_{2,3,4}$ account for the second, the third and the fourth-order chromatic dispersion, respectively. $\tau $ is the fast time in the reference frame moving with the group velocity of the light. $i|E|^2E$ corresponds to nonlinearity described only by Kerr effect because we consider that the pulse width is larger than  $1 ps$. In this case we can neglect the Raman scattering. $\Delta$ is the cavity detuning.
The model Eq. (\ref{eq:E}) is valid in the limit of a high cavity finesse,  the nonlinear phase shift, and losses have to be smaller than unity.  Finally, we assume that the optical field maintains its polarization as it propagates along the fiber. Note that the LL equation without higher order dispersions has a broad applicability that the fiber or spatial resonator. It has been shown that, the LL model could describe the  Kerr-comb evolution in whispering-gallery-mode resonators, where the $t$ is the time, and the variable $\tau $ is the azimuthal angle \cite{Chembo10,Chembo13}. Indeed, they show that at low-threshold, wide-span combs can emerge as the well known  temporal localized structures reported in  \cite{Scroggie}.

The homogeneous steady states (HSS) of Eq. (\ref{eq:E}) satisfy $S=[1+i(\Delta -\left| E_{s}\right| ^{2})]E _{s}$. Obviously, high order dispersion do not affect these solutions, they are thus  identical to the ones of LL model \cite{LL}. The  linear stability analysis of the HSS with respect to finite frequency perturbations of the form  $\exp (\lambda t-i\Omega \tau)$ yields the eigenvalue
\begin{eqnarray}
\lambda=-1+iB_3\Omega^3\pm \sqrt{I_s^2-(-\Delta+2I_s+B_2\Omega^2+B_4\Omega^4)^2} \nonumber
\end{eqnarray}
where $I_s=|E_s|^2$ corresponds to the uniform intensity background of light. The thresholds associated with modulational instability are $I_{m1}=1$ and  $I_{m2}=[2\kappa+\sqrt{\kappa^2-3}]/3$ with $\kappa= B_2^2/(4 B_4)+\Delta$. In the monostable case ($\Delta<\sqrt{3}$),  the primary instability threshold is degenerated: Two critical frequencies appear spontaneously and simultaneously $\Omega^2_{l}$ and $\Omega^2_{u}$ where $\Omega^2_{l,u}=[-B_2\pm\sqrt{B^2_2+4B_4(\Delta-2)}]/(2B_4).  $ At the threshold $I_{m2}$, a new, large, critical frequency appear $\Omega^2=-B_2/2B_4$. The existence of two thresholds associated with the modulational instability allows the instability domain to be bounded as shown in Figure 1.

The linear stability analysis shows that the third order dispersion affects neither the threshold nor the frequency associated with the periodic train of pulses. This analysis provides a linear velocity for temporal dissipative structures. The linear velocity reads
\begin{equation}
V_l=\frac{\partial Im(\lambda) }{\partial \Omega}=\frac{3B_3\sqrt{\left( 2-\Delta\right) \,B_4}}{B_4}
\label{linvelo}
\end{equation}
This simple expression indicates that in the absence of the third order dispersion $B_3=0$, the trains of temporal pulses are motionless $V_l=0$. The third order dispersion pushes the train of temporal pulses to move along the $\tau$ direction as we shall see in the next section.

\section{Weakly nonlinear analysis}

In this section, we focus on regime where the first bifurcation is not degenerate where both frequencies coincides, i.e., $\Omega^2_{l}=\Omega^2_{u}$. This condition yields $B_2^2+4(\Delta-2)B_4=0$. In this case the unstable zone is delimited by two  modulational instability thresholds $Y_{1m}=1$ and $Y_{2m}=5/3$ and the critical frequency at both bifurcation points is $\Omega_c^4= ( 2-\Delta)/B_4$. In this case, we have 
\begin{equation}
\Delta=2-(B_2^2/4B_4) \,\,\, {\text {and}} \,\,\,  \Omega_c^4= B_2^2/(4B_4^{2})
\end{equation}
To evaluate nonlinear solutions that emerge from the first threshold of the modulational instability, we use a weakly nonlinear analysis. To this end we decompose the electric field into its real and imaginary parts as $E=X_1+iX_2$. and we introduce the deviation $X_{1,2}=X_{1s,2s}+u(t,\tau)$. We explore the vicinity of the first threshold associated with the modulational instability. We choose a small parameter  $\epsilon^2=S-S_{1m}$ which measures the distance from the modulational instability threshold. Next, we expand in powers of $\epsilon$, the input field $S$, the variables $u_1$ and $u_2$ and the homogeneous stationary solutions $X_{1s}$ and $X_{2s}$ 
\begin{eqnarray}
S&=&S_{1m}+\epsilon p_1+\epsilon^2 p_2+...\\
u_{1,2}&=&\epsilon  u_{10,20}+\epsilon^2  u_{11,12}+\epsilon^3  u_{21,22}+...\\
X_{1s}&=&X_{1m}+\epsilon a_1+\epsilon^2 a_2+...\\
X_{2s}&=&X_{2m}+\epsilon b_1+\epsilon^2 b_2+...
\end{eqnarray}
where $X_{1m}=1/S_{1m}$ and $X_{2m}=(1-\Delta)/S_{1m}$ are the values of the real and the imaginary parts at the threshold associated with the modulational instability with $S_{1m}^2=[1+(\Delta -1)^2]$. We also introduce a slow time $T=\epsilon^2 t$. The solutions at the leading order in $\epsilon$ are
\begin{equation}
u_{10,20}=[1, (2-\Delta)/\Delta]A_1\exp{i(\Omega_c\tau+\phi T)+C.C.} 
\end{equation}
where $[1, (2-\Delta)/\Delta]$  is the eigenvector of the linearized operator at the modulational instability point,
C.C. denotes the complex conjugate, and $\phi$ is the phase. The solvability condition at the leading order imposes that $p_1=0$. 

The solvability condition at the third order of $\epsilon$ yields the following amplitude equation for the wave number of the fastest growing frequency is:
\begin{eqnarray}
\frac{1}{2S^2_{1m}}\frac{\partial A}{\partial t}=\frac{S-S_{1m}}{S_{1m}(2-\Delta)^2}A-\ ({f}_{1}\left( \Delta\right)+ i\,{f}_{2}\left( \Delta\right))\left|A \right| ^{2}A 
\label{AmpEq}
\end{eqnarray}
Where 
\begin{equation}
f_{1}\left(\Delta\right)=\frac{ac+bd}{c^2+d^2} \, \, {\text {and}   }\, \, f_2\left(\Delta\right)=\frac{bc-ad}{c^2+d^2}
\label{real} 
\end{equation}
With
\[a=2\,\left( 200\,{B}_{3}^{2}\,\left( 2\,\Delta-3\right) \,{\Omega}^{6}-{\left( \Delta-2\right) }^{2}\,\left( 342\,\Delta-521\right) \right) \]
\[b=20\,{B}_{3}\,\left( {\Delta}^{2}-12\,\Delta+16\right) \,{\Omega}^{3}\]
\[c=-{\left( \Delta-2\right) }^{2}\,{\Delta}^{2}\,\left( 10\,{B}_{3}\,{\Omega}^{3}-9\,\Delta+18\right) \,\left( 10\,{B}_{3}\,{\Omega}^{3}+9\,\Delta-18\right) \]
\[d=20\,{B}_{3}\,{\left( \Delta-2\right) }^{2}\,{\Delta}^{2}\,{\Omega}^{3}\]
The amplitude equation (\ref{AmpEq}) is expressed in term of the amplitude $A=\epsilon A_1$.
For a given values of $B_3$ and $B_4$ such that  $f_1(\Delta)>0$, for any input field  $S>S_{1m}$, the nonlinear solutions are stable, and this bifurcation is called supercritical. While, if  $f_1(\Delta)<0$, the nonlinear solutions bifurcate subcritically and are unstable when  $S<S_{1m}$. 

\begin{figure}[bbp]
\begin{center}
\includegraphics[width=7cm,height=12cm]{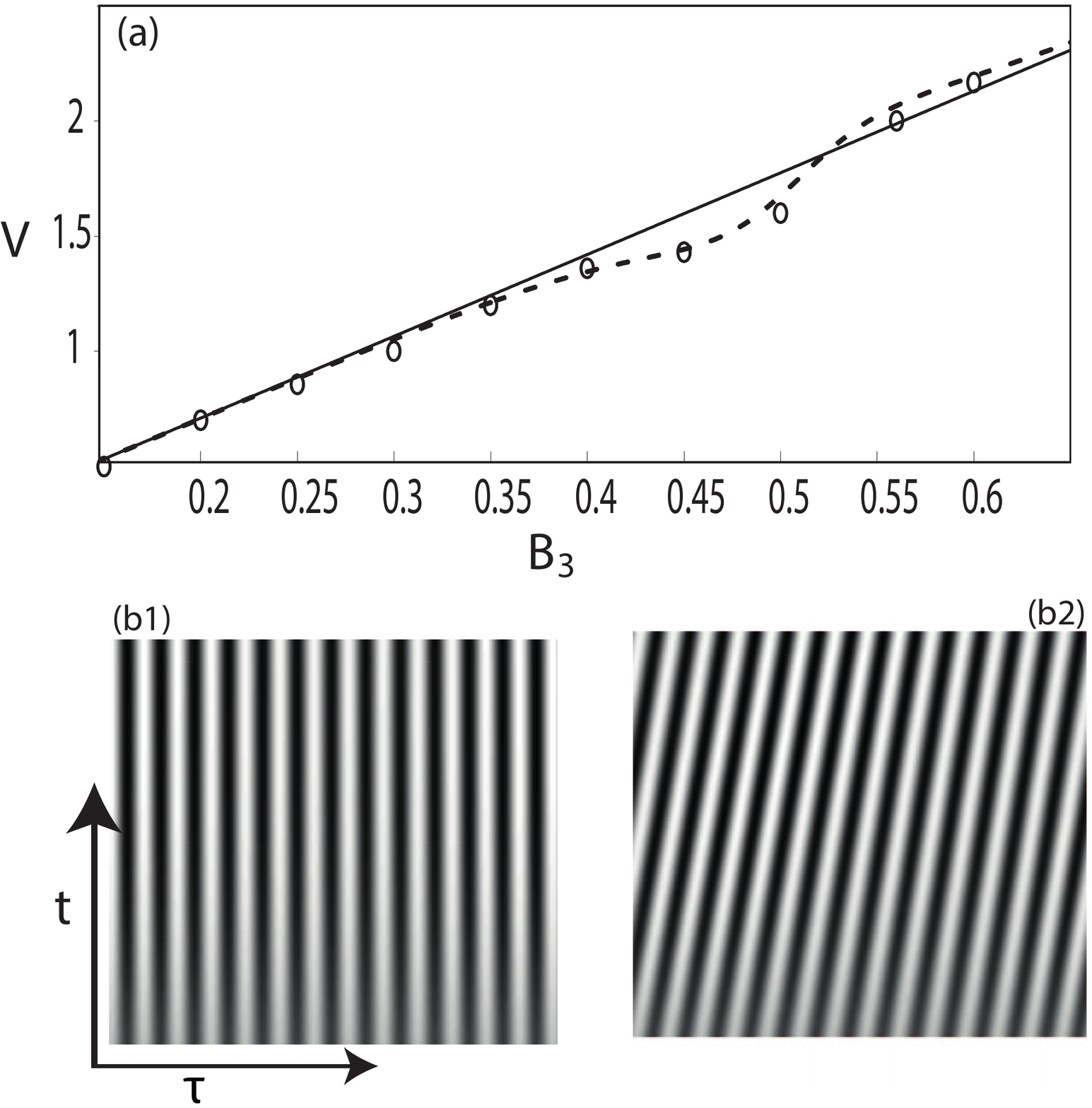}
\end{center}
\caption{(a) Variation of the velocity of trains of short pulses with respect to $B_3$. The solid curve denotes the linear velocity and the dashed curve represents the velocity that includes the nonlinear correction for a fixed value of the injected beam. The velocity obtained from numerical simulations of Eq. (1) is denoted by circles. The $\tau$-$t$ maps showing the time evolution of periodic structures that emerge from supercritical bifurcation for $(b_1)$ $B_3=0$ and  $(b_2)$ $B_3=0.12$. Other parameters are  $S=1.05$, $\Delta=1.3$, $B_4=0.5$, $B_2=-1.1832$}
\label{Fig2}
\end{figure}
Next, the nonlinear analysis allows  to calculate the amplitude and the phase of periodic solutions that emerge from supercritical bifurcation. Assuming that $A=A_1\exp {i(\phi T+\Omega\tau)}$,  we obtain
\begin{eqnarray}
\left|A_1 \right|^{2}=\frac{1}{f_1(\Delta)}\frac{\epsilon^{2}}{S_{1m}(2-\Delta)^2} \nonumber
\end{eqnarray}
\begin{eqnarray}
\phi=-2S^{2}_{1m}f_2(\Delta)\left|A_1 \right|^{2} \nonumber
\end{eqnarray}

According to these results, it is obvious that in addition to the parameters of the system ($S$, $B_2$,$B_3$, $B_4$ and $\Delta$), the distance from the modulational instability threshold plays an important role in the dynamics of periodic solutions. Both amplitude $\left|A_1\right|^{2}$ and the phase $\phi$ are proportional to this distance.
The nonlinear phase $\phi$ is caused by the third order dispersion. When taking into account the nonlinear correction, the velocity takes the following form:
\begin{equation}
v=\frac{3B_3\sqrt{\left(2-\Delta\right)B_4}}{B_4}+\frac{\partial \phi}{\partial\Omega}
\end{equation}
\begin{equation}
v=\frac{3B_3\sqrt{\left(2-\Delta\right) B_4}}{B_4}-h(\Delta,B_4)B_3 (S-S_{1m})\nonumber
\end{equation}
With
\begin{equation}
h=-\frac{\partial \left(\frac{f_2}{f_1}\right)}{\partial \Omega} \frac{2S_{1m}}{\left( 2-\Delta\right)^2}
\end{equation}

$h$ is a the velocity correction function depending on $\Delta$ and $B_4$ and proportional to $B_3$ and $S$. The velocity of moving periodic solutions as a function of the third order dispersion coefficient is shown in Fig. (2.a). The linear velocity Eq (\ref{linvelo}) of the periodic structures is affected by the third order dispersion. This nonlinear correction can increase or decrease the velocity depending on the values of $B_3$ and the distance from threshold of instability. To check this result, we numerically integrate Eq. (1) with periodic boundary conditions. The numerical results are plotted together with the analytical expression of the velocity. The comparison between the numerical results and the analytical ones agree as shown in Fig. (2.a). The $\tau$-$t$  map of the Fig. (2b1) describes the time evolution of periodic structures in the absence of the thirds order dispersion. When third order dispersion is neglected, the structure is always stationary. However, in the presence of the third order dispersion, the temporal structures that propagates inside the cavity undergo a drift from a round cavity trip to another with a well defined velocity as shown in the  $\tau$-$t$  map of the Fig. (2b2).

\section{Subcritical modulational instability and temporal localized structures}

Localized structures are usually excited in the pinning region involving the homogeneous steady state and the periodic dissipative structures \cite{Scroggie,TML94}. Therefore, the occurrence of a subcritical modulational instability is often the prerequisite condition for the emergence of TLSs. By now, a large body of literature exists on the study of localized structures in biology, chemistry, physics, and mathematics (see some of overviews on this issue \cite{Mandel_2004,Rosanov, Marcel_REsidori}). This field is now attracting growing interest in optics because of its potential application in information technology. In particular, they could be used for all-optical storage, all-optical reshaping and wavelength conversion \cite{LEO}. The aim of this section is twofold. Firstly, to determine through a weakly nonlinear analysis performed in Sec. 3, the threshold associated with the formation of bright TLS in the absence of a third order dispersion. Secondly, to study the role of the of the third order dispersion that breaks the reflexion symmetry ($\tau\rightarrow-\tau)$, and leads to the formation of moving TLS. Equation (\ref{eq:E})  admits a variety of temporal  TLSs \cite{Tlidi-Gelens,Tlidi-Lyes}. These solutions exhibit a complex homoclinic snaking type of bifurcation as shown in Ref. \cite{Tlidi-Gelens}. This means that, the system exhibits a high degree of multistability in a finite range of parameters often called the pinning region. There exist an infinite number of stable TLSs each of them characterized either by an odd number  or even number of peaks or dips. The configuration that maximizes the number of peaks or dips in the pattern corresponds to trains of short pulses with well defined repetition rate. An example of  motionless  TLS  having a single peak and a single dip is displayed in the $\tau$-$t$ map of the Fig. 3(a1). This figure is obtained for $B_3=0$. This solution is symmetric as shown in the cross section  along the $\tau$ coordinate [Cf. Fig.  3(b1)]. When taking into account the third order dispersion $B_3 \neq 0$, a single peak TLS exhibits a spontaneous motion as shown in the $\tau$-$t$ map of Fig. 3(a2). The cross section along the $\tau$ coordinate shows an asymmetry in the intensity profile of the intracavity field  [see Fig.  3(b2)]. 
\begin{figure}[bbp]
\begin{center}
\includegraphics[width=12.5cm,height=12cm]{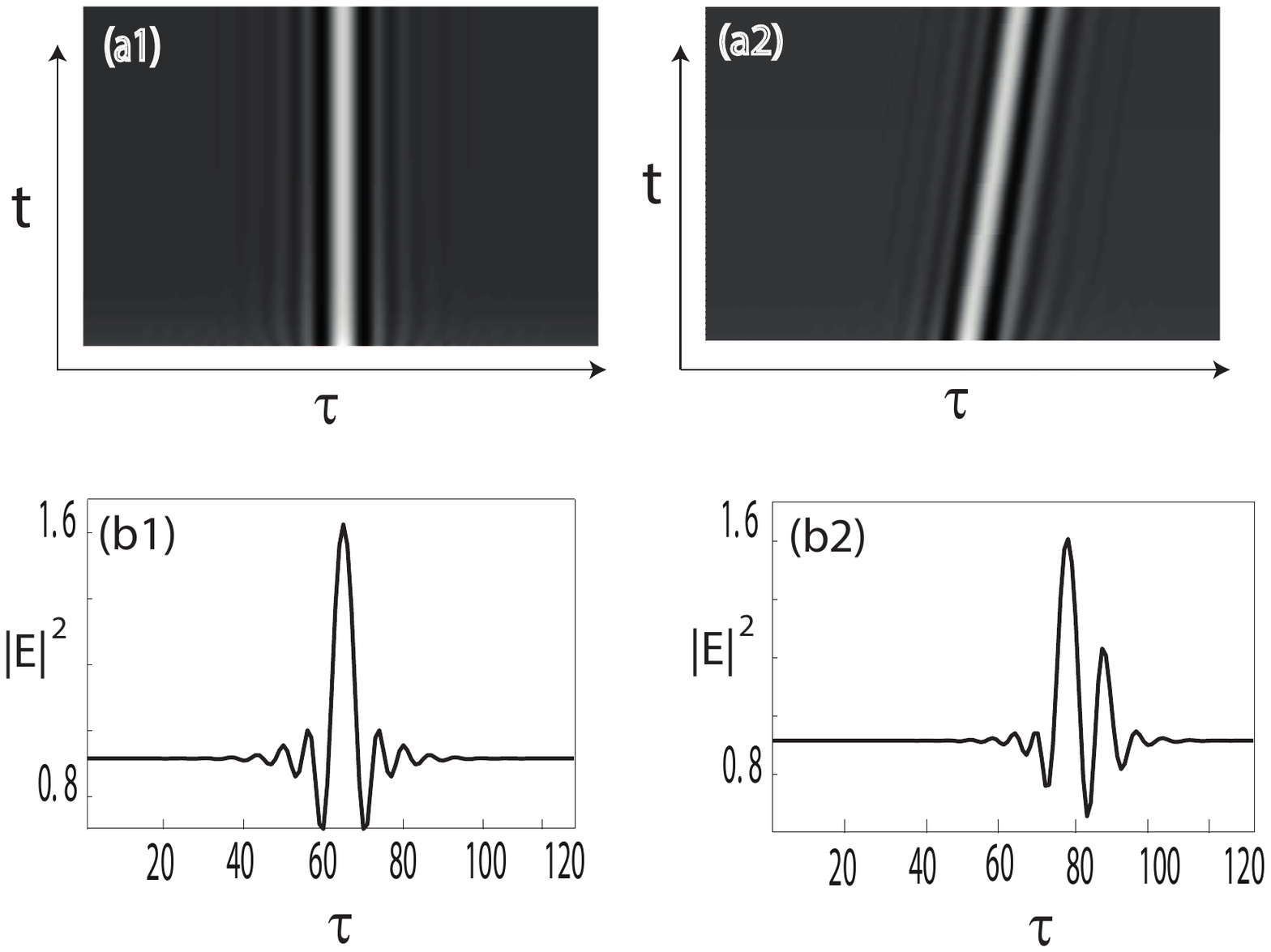}
\end{center}
\caption{$(a_1,a_2)$, $\tau$-$t$ map showing the time evolution of bright TLS that emerge from subcritical bifurcation, $(a_1)$ $B_3=0$, $(a_2)$ $B_3=0.12$. $(b_1)$ Stationary  bright TLS, $B_3=0$, $(b_2)$Single moving bright soliton, $B_3=0.12$. Parameters are $B_2=-0.7483$, $B_4=0.5$, $\Delta=1.72$ and $S=1.228$.}
\label{Fig3}
\end{figure}
\begin{figure}[bbp]
\begin{center}
\includegraphics[width=18cm,height=10cm]{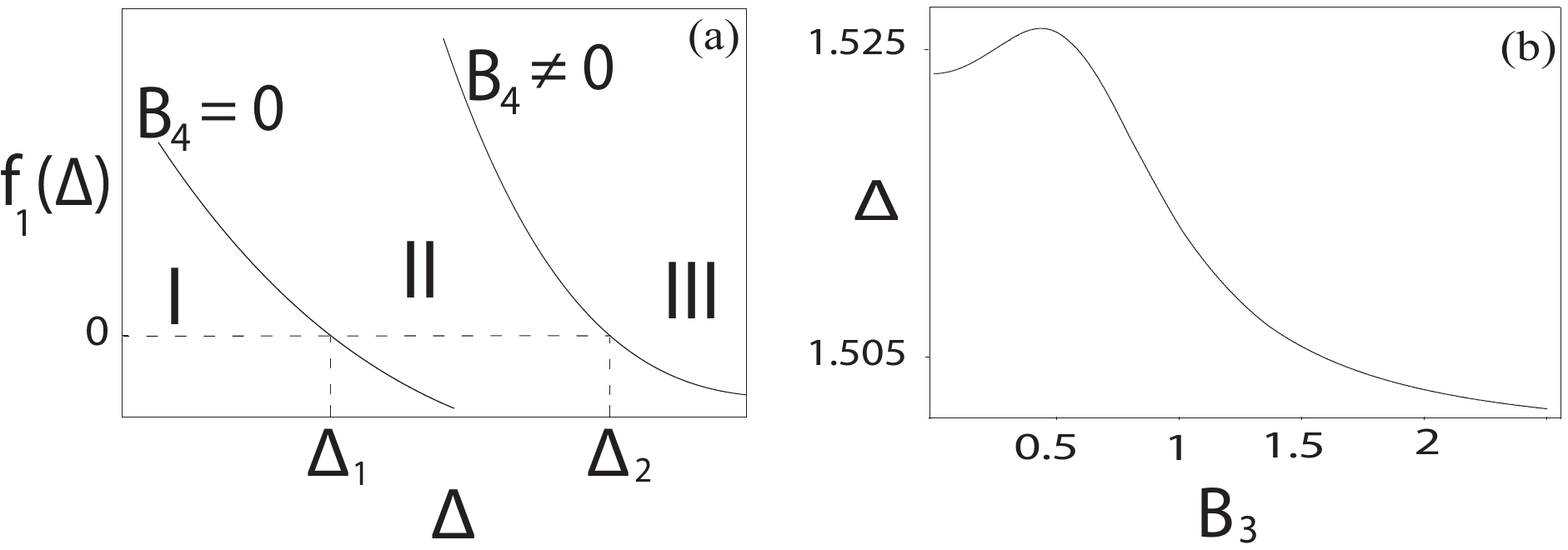}
\end{center}
\caption{(a)Variation of $f_1(\Delta)$ as function of $\Delta$,the threshold is shifted from $\Delta_1 = 1.367$ when $B_4=0$ to $\Delta_2 = 1.523$ when $B_4\neq0$, $B_3=0$.(b) Variation of $\Delta$ threshold $(f_1(\Delta)=0)$ according to $B_3$,  $B_4\neq0$}
\label{Fig4}
\end{figure}
\begin{figure}[bbp]
\begin{center}
\includegraphics[width=18cm,height=10cm]{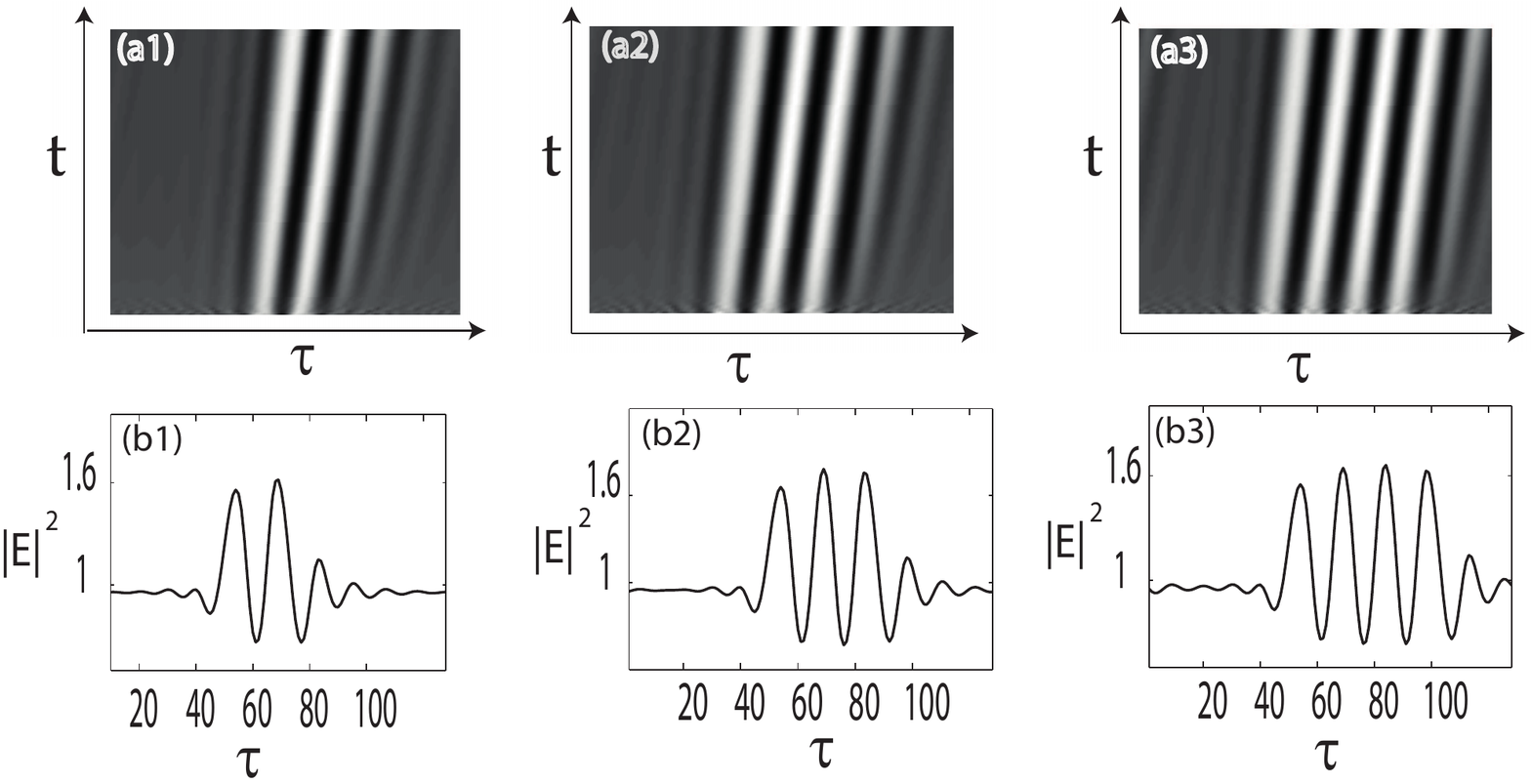}
\end{center}
\caption{Moving multi-peaks localized structures $(b_1)$ two $(b_2)$ and $(b_3)$ four peaks. They are asymmetric solutions since the third order dispersion breaks the reflexion symmetry ($\tau\rightarrow-\tau)$.  $(a_1) $, $(a_2)$ and $(a_3)$ are $\tau$-$t$ maps showing the time evolution of bright moving solitons associated with $(b_1)$, $(b_2)$, and $(b_3)$, respectively. They are obtained numerically by integrating eq (\ref{eq:E}) with periodic boundary conditions. Parameters are same as Figure 3}
\label{Fig5}
\end{figure}
They are obtained by numerical integration of (\ref{eq:E})  with periodic boundary conditions.  

The weakly nonlinear theory presented in Sec. 3, cannot describe temporal localized structures because it does not take into
account the nonadiabatic effects that involve the fast temporal scales which are responsible for the
stabilization of TLS \cite{Pomeau}. The inclusion of amended terms in amplitude equations can
capture this dynamics \cite{C1,C2}. However, the weakly nonlinear analysis provides an information
about the threshold associated with the appearance of bright temporal localized structures.  
Indeed, the sign of $f_1(\Delta)$ provides an information about the nature of the bifurcation that
can exist in this resonator. An explicit expression for $f_1(\Delta)$ is given by the equation (\ref{real}). When $f_1(\Delta)> 0$, the bifurcation is supercritical and the train of periodic solutions emerge beyond the  modulational instability threshold. However, when ($f_1(\Delta)< 0$) the bifurcation is subcritical and periodic solutions can exist below the  modulational instability threshold. In this case, there exists an hysteresis loop involving a coexistence between the homogenous steady state and the periodic solution which are both linerarly stable. In this region, there exists a pinning zone for which temporal localized structures are stable. The condition ($f_1(\Delta)= 0$) gives the threshold of the appearance of TLS. The plot of the function  $f_1(\Delta)$ shows that even when $B_3=0$, the threshold associated with the formation of TLS is shifted with respect to the detuning parameter, i.e., $\Delta= $521/342$\approx 1.523$, as shown in Fig. 4(a). Indeed, when $B_4=B_3=0$, we recover the classical condition of the inversion of the bifurcation derived by Lugiato and Lefever $\Delta=41/30\approx 1.367$ \cite{LL}. For $B_3=0$ and $B_4=0$, transition from zone I to zone II indicates the change in the nature of the bifurcation, i.e., from super- to sub- critical modulational instability [see  Fig. 4(a)]. When $B_3=0$ and $B_4\neq0$, transition from super- to sub- critical Modulational instability occurs between zone II and zone III as shown in Fig. 4(a). When taking into account the third and the fourth order dispersions ($B_3\neq0$ and $B_4\neq0$), the real part of the coefficient of the nonlinear term in the amplitude equation  $f_1$ depends on the third order dispersion. In the Fig. 4(b) we plot the function $f_1=0$ in the plane ($\Delta$, $B_3$). The solid line in this figure indicates the threshold associated with the appearance of TLS. We can then see from Fig. 3 and Fig. 4 that higher order dispersion induced a spontaneous symmetry breaking instability and allows bright TLSs to appear for larger intensity of the injected beam.

It has been shown that dark temporal localized structures exhibit a homoclinic snaking type of instability \cite{Tlidi-Gelens}. In the rest of this paper we will show that the same type of behavior occurs for moving bright TLS when taking into account of both third and fourth orders of dispersion ($B_3\neq0$ and $B_4\neq0$).
The homoclinic nature of these solutions implies that for a given set of control parameters,
the number and the temporal distribution of both bright and dark TLS immersed in the bulk of the
homogeneous steady state are determined only by the initial condition. Temporal localized structures may, therefore, be used
for signal processing since the addition or the removal of a TLS simply means the change from one
solution to another. Note that the same model equation (\ref{eq:E})  using fourth-order diffraction instead
of dispersion has previously also been proven to support higher-order spatial effects on bright
spatial solitons \cite{Gelens07,Kockaert}. Moving temporal localized structures involving multipeak solutions are shown in Fig. 5. They are obtained for the parameter values as the single peak TLS of the Fig. 3. 
\section{Conclusion}
In conclusion, we have studied the impact of the effects of high orders of dispersion on the dynamics of temporal localized structures in photonic crystal fiber resonator pumped by a continuous wave. Both bright and dark temporal localized structures are possible. Without fourth order dispersion dark localized structures do not exist. They consist of asymmetric moving peaks or dips in a uniform background of the intensity profile. The number of  moving localized peaks structures and their temporal distribution is determined solely by the initial conditions.  We have focused the analysis on bright temporal localized structures. We have characterized this motion by computing the velocity of bright temporal localized structures.  The weakly nonlinear analysis in the vicinity of the first threshold associated with the modulational instability is performed. This analysis first shows that the threshold associated with the temporal localized structures is shifted from   $\Delta=41/30\approx 1.367$ to $\Delta= $521/342$\approx 1.523$. Second, the weakly nonlinear analysis allows to estimate the linear and the nonlinear velocity associated with the moving temporal localized structures. Numerical simulations of the governing model for all fiber photonic crystal resonator are performed. Numerical solutions are in close agreement with the analytical predictions.  Our study confirms the possibility of reducing the size of temporal localized structures close to the zero dispersion wavelength by using photonic crystal fibers. 

\section*{Acknowledgment}
M .T received support from the Fonds National de la Recherche Scientifique (Belgium). This research was supported by the Interuniversity Attraction Poles program of the Belgian Science Policy Office, under grant IAP 7-35.


\end{document}